\documentclass{article} 
\usepackage{graphicx,floatflt,amssymb} 
\def\mpl{\ifmmode \overline M_{P}\else $\overline M_{P}$\fi} 
\begin{document} 
\renewcommand{\thefootnote}{\fnsymbol{footnote}} 
\begin{flushright} 
SINP/TNP/02-04\\ 
hep-ph/0202147 
\end{flushright} 
\begin{center} 
{\Large \bf  Power law enhancement of neutrino mixing angles \\ in 
extra dimensions}\\ 
\vspace*{1cm} 
{\large\sf Gautam Bhattacharyya ${}^1$, Srubabati Goswami ${}^1$, 
          Amitava Raychaudhuri ${}^2$} \\ 
\vspace{10pt} 
{\small 1. Saha Institute of Nuclear Physics, 1/AF Bidhan 
        Nagar, Kolkata 700064, India \\ 
     2. Department of Physics, University of Calcutta, 92 Acharya 
        Prafulla Chandra Road, \\ Kolkata 700009, India } 
 
\normalsize 
\end{center} 
 
\begin{abstract} 
We study the renormalization of the $llHH$-type Majorana neutrino 
mass operator in a scenario where there is a compactified extra 
dimension and the fields involved correspond to only the standard 
model particles and their Kaluza-Klein excitations. We observe 
that in a two flavour scenario, where one of the neutrinos is 
necessarily $\nu_\tau$, it is indeed possible to generate a large 
mixing at $\sim$ 100 GeV starting from a very small mixing near 
the ultra-violet cutoff $\sim$ 30 TeV. {\em En passant}, we also 
derive the Higgs mass upper and lower limits from perturbative 
unitarity and stability of the potential, respectively. 
 
\vskip 5pt \noindent 
\texttt{PACS Nos:} 14.60.Pq, 11.10.Kk, 11.30. Hv., 11.25.Mj  
 
\end{abstract} 
\vskip 20pt 
 
\setcounter{footnote}{0} 
\renewcommand{\thefootnote}{\arabic{footnote}} 
 
If lepton number is not a good symmetry of the Lagrangian, then, 
without enlarging the standard model (SM) particle content, a neutrino 
Majorana mass operator can be written as (with $i,j$ as generation 
indices)  
\begin{eqnarray} 
-{\cal{L}}^{\rm SM} = {{\kappa_{ij}}\over{M_X}} \bar{l^c}_i l_j H H + 
{\rm h.c.} 
\label{smlag} 
\end{eqnarray} 
This dimension-5 operator can be viewed as a consequence of 
integrating out a superheavy right-handed neutrino of mass $\sim M_X$ 
which is exchanged at the tree level. Here $l$ is the SM lepton 
doublet and $H$ is the SM Higgs doublet. Eq.~(\ref{smlag}) gives a 
neutrino mass matrix $m_{ij} \sim \kappa_{ij} (v^2/M_X)$, where $v$ is 
the vacuum expectation value of the SM Higgs.  Assuming $\kappa \sim 
1$, a choice of $M_X \sim 10^{15}$ GeV produces $m \sim 0.1$ eV.  It 
has been pointed out in \cite{babu1,chan,rest} that starting from a 
small mixing angle between two active neutrinos at a high scale, a 
large mixing between them can be generated at a low scale due to 
renormalization group (RG) evolution. In this paper we intend to 
investigate the renormalization of the above operator in 
extra-dimensional models.  For simplicity, we consider only one 
additional space dimension which is compactified on a circle. Since 
both solar and atmospheric neutrino data prefer large neutrino mixing, 
our primary aim is to examine whether the extra-dimensional models can 
reproduce this feature. We restrict ourselves only to the case of 
oscillation between two active generations where one of the two 
neutrinos is necessarily $\nu_\tau$. Even though the mass scales in 
such models are expected to be quite close -- around 1 TeV in our 
choice -- and the energy range for RG running small, we will show that 
because of the power law evolution of the $\kappa$ operator, the 
neutrino mixing angle runs rather fast once the Kaluza-Klein (KK) 
modes of the higher dimensional fields open up. As a result, even if 
the two-flavour mixing angle happens to be quite small near the 
ultra-violet cut-off $\Lambda \sim {\cal{O}}$ (10 TeV), where the 
textures are defined, near-maximal mixing can be generated at the 100 
GeV scale. If the mixing is large at the high scale then it undersgoes 
further enhancement due to RG running. 
 
We stick to a very simple extra-dimensional scenario in which the extra 
space dimension ($y$) is compactified on a circle of radius $R$, i.e., $y 
\leftrightarrow y + 2\pi R$.  In our simple approach, all fermions are 
localized at the brane at $y = 0$, but the bosons can also travel in the 
bulk \cite{anto,ddg,addj}. In the effective 4-dimensional 
representation, after the fifth coordinate is integrated out, the 
resulting Majorana mass operator looks like 
\begin{eqnarray} 
-{\cal{L}}^{\rm eff} = {{\kappa_{ij}}\over{\pi M^2 R}} \bar{l^c}_i 
l_j H_0 H_0 + {\rm h.c.} 
\label{efflag} 
\end{eqnarray} 
Above, $H_0(x)$ is the zero mode of the KK excitations of the doublet 
scalar in five dimensions: $H(x,y) = (1/\sqrt{\pi R}) 
\sum_{n=-\infty}^{\infty} H_n(x) ~{\rm exp}~(iny/R)$. $M$ corresponds 
to some higher dimensional mass scale beyond which new physics sets 
in. 
 
The neutrino mass matrix is given by $m_{ij} \sim \kappa_{ij} (v^2/\pi 
M^2 R)$. For definiteness, we assume $\mu_0 \equiv R^{-1} = {\cal O}$ (1 
TeV). $\mu_0$ determines the mass splittings of the KK excitations. The 
appearance of $M$ may be interpreted as a consequence of integrating out 
some physical states around $\sim M$ (e.g., a right-handed neutrino $N$ 
with a mass $M$ that couples like $LHN$) which leads to the effective 
operator in Eq.~(\ref{efflag}). Thus below the scale $M$ the theory is 
essentially non-renormalizable in the sense that a heavy state is 
integerated out leading to effective Lagrangian in Eq.~(\ref{efflag}). 
Since we are basically interested in the evolution of $\kappa$, which in 
turn requires the running of gauge, Yukawa, and Higgs self-couplings, it 
seems quite reasonable to associate the cut-off parameter $\Lambda$ with 
$M$.   
 
Now we attempt to briefly address the issue of a second kind of 
non-renormalisibility which stems mainly from the presence of an {\em 
infinite} tower of KK states after compactification to 4 dimensions (for 
an extensive discussion see Appendix B of \cite{ddg}). In fact, as 
stressed in \cite{ddg}, the couplings do not strictly run in a 
non-renormalizable theory. Instead they receive finite quantum 
corrections whose magnitudes explicitly depend upon the cut-off $\Lambda 
\sim M$. It also turns out that very often the mathematical dependence 
of a coupling on $\Lambda$ is identical to its scale-dependence that 
follows from a naive calculation assuming a renormalizable theory. Since 
the root of this non-renormalisibility lies in having an {\em infinite} 
KK tower, the remedy, as suggested in \cite{ddg}, is to consider a {\em 
truncated} KK series which has been shown to serve as an excellent 
approximation for calculating the scale dependence of couplings.  Under 
the above guideline, we continue to describe the quantum corrections of 
couplings as their RG running. Indeed, all the couplings have to remain 
perturbative throughout the energy interval $M_Z <\mu<M$, and a rough 
estimate of the hierarchy \cite{bando}, namely $(M/\mu_0)^\delta \sim 
\ln~(M_{\rm GUT}/M_W)$, with $M_{\rm GUT}$ as the 4-dimensional GUT 
scale and $\delta$ as the number of extra dimensions, yields $M \sim 30 
\mu_0$ for $\delta = 1$.

Here, for the sake of clarity, we stress that $M$ should not, in
general, be equated to the 5-dimensional Planck scale $M_\star$. In
fact, it follows from the relation $M_P^2 = M_\star^2 (M_\star
R)^\delta$, where $M_P$ is the 4-dimensional Planck scale, that
$M_\star \sim 10^{10} - 10^{11}$ TeV for $\delta = 1$ and $R^{-1} = 1$
TeV.  Thus $M_\star \gg M$ and hence quantum gravity effects on the
effective Majorana mass operator at the scale $M$ or below are
insignificant.
 
Assuming $\kappa \sim 1$, a further suppression of 9 orders of 
magnitude is required to produce a neutrino mass of order 0.1 eV. Such 
a suppression may come from a distant brane where lepton number ($L$) 
is violated and the effect at the brane under consideration is damped 
by the distance between the two branes \cite{addj}.  
 
Since quark mixing angles are small, our main curiosity in this paper 
will be to check whether a small $\nu_\tau$-$\nu_e$ or 
$\nu_\tau$-$\nu_\mu$ mixing near $\Lambda \sim M$ can indeed become 
large at $M_Z$ due to power law RG running. The mixing angle depends 
not on the absolute value of $\kappa_{ij}$, but on the degree of 
degeneracy of $\kappa_{11}$ and $\kappa_{22}$. We will need to tune 
this difference at $\sim M$ to obtain a large mixing angle at $\sim 
M_Z$. In fact, we have found that this tuning ensures the mixing at 
$M_Z$ to be large for essentially {\em any} initial mixing, small or 
large.

The presence of extra dimension modifies the running of $\kappa$ 
(matrix) in the region $\mu > \mu_0$ as follows: 
\begin{eqnarray} 
\label{kappa} 
16\pi^2 \frac{d\kappa}{d\ln\mu} & = & \left(-3g_2^2  + 
2\lambda + 2S\right)t_\delta \kappa - \frac{3}{2} 
t_\delta\left[\kappa(Y_l^\dagger Y_l) + (Y_l^\dagger 
Y_l)\kappa\right],  
\end{eqnarray} 
where $S = {\rm Tr}~ (3Y_u^\dagger Y_u + 3 Y_d^\dagger Y_d + 
Y_l^\dagger Y_l)$, and $t_\delta = (\mu/\mu_0)^\delta X_\delta$.  In 
Eq.~(\ref{kappa}), $t_\delta$ controls the power law behaviour, where 
$X_\delta$ can be expressed in terms of the Euler Gamma function as 
$X_\delta = 2 \pi^{\delta/2}/\delta \; \Gamma(\delta/2)$.  For $\delta 
= 0(1)$, $X_\delta = 1(2)$. It is important, for later discussions, to 
bear in mind that Eq.~(\ref{kappa}) is homogenous in $\kappa$. 
 
The running of the Yukawa couplings ($Y_u, Y_d$) and Higgs 
self-coupling ($\lambda$) for $\mu > \mu_0$ are given by 
\begin{eqnarray} 
\label{yukawa} 
16\pi^2 \frac{d Y_u}{d\ln\mu} & = & \frac{3}{2} 
t_\delta\left(Y_u Y_u^\dagger Y_u 
- Y_d^\dagger Y_d Y_u\right) + t_\delta S Y_u - t_\delta \left(8g_3^2 + 
\frac{17}{20} g_1^2 + \frac{9}{4} g_2^2\right) Y_u, \nonumber\\ 
16\pi^2 \frac{d Y_d}{d\ln\mu} & = & \frac{3}{2} 
t_\delta\left(Y_d Y_d^\dagger Y_d 
- Y_d Y_u Y_u^\dagger \right) + t_\delta S Y_d - t_\delta \left(8g_3^2 + 
\frac{1}{4} g_1^2 + \frac{9}{4} g_2^2\right) Y_d, \\ 
16\pi^2 \frac{d\lambda}{d\ln\mu} & = & 12 t_\delta \lambda^2 - 
\left(\frac{9}{5} g_1^2 + 9 g_2^2\right)t_\delta \lambda + 
\frac{9}{4} t_\delta 
\left(\frac{3}{25} g_1^4 + \frac{2}{5} g_1^2 g_2^2 + g_2^4\right) 
\nonumber \\ 
& + & 4 S \lambda - 
4 {\rm Tr}~\left[(Y_l^\dagger Y_l)^2 + 3 (Y_d^\dagger Y_d)^2 
+ 3 (Y_u^\dagger Y_u)^2\right] ~. \nonumber 
\end{eqnarray} 
It should be noted that in the limit $\delta = 0$ (i.e., $t_\delta = 
1$) one reproduces the SM expressions \cite{babu1,chan,drees,mv} 
which control the evolution in the interval $M_Z < \mu <\mu_0$. 
We stress here that our calculation of $\kappa$ evolution agrees with 
that of \cite{drees} who have pointed out a small error in the 
original calculations of \cite{babu1,chan}: more specifically, 
the numerical factor in front of the leptonic Yukawa contribution in 
Eq.~(\ref{kappa}) is indeed 3/2 rather than 1/2.  
 
The evolution of the gauge couplings in an extra-dimensional 
scenario have been worked out in 
\cite{ddg}, and for $\mu > \mu_0$ are given by 
\begin{eqnarray} 
\label{gauge} 
16\pi^2 \frac{d g_i}{d\ln\mu} & = & b_i g_i^3, ~~~{\rm where} \\ 
b_1 & = & 41/10 + (t_\delta - 1)(1/10), \nonumber \\ 
b_2 & = & -(19/6) - (t_\delta - 1)(41/6), \nonumber \\ 
b_3 & = & -7 - (t_\delta - 1)(21/2). \nonumber 
\end{eqnarray} 
In the interval $M_Z < \mu <\mu_0$, the gauge couplings run as in the 
SM and the corresponding beta functions are obtained by putting 
$t_\delta = 1$ in Eq.~(\ref{gauge}). 
 
The computational procedure behind the power law running behaviour is 
simple \cite{ddg,abelking}. In the scenario under consideration, gauge 
bosons and scalars have KK excitations, but fermions are localised at a 
brane, which is a fixed point. The external boson legs in any diagram 
are their KK zero modes which represent their SM states. In the loop 
diagrams there can be either one or two internal KK modes. If there is 
only one, then each time a KK threshold is crossed, the diagram 
contributes the same as in the SM regardless of the KK number of the 
internal line. Such a situation may arise only when an internal boson 
meets a fermion at the brane where KK number is not conserved due to the 
breakdown of the fifth-dimensional translational invariance at the fixed 
point. If there are two internal KK modes, then both should have the 
same KK number as the latter is assumed to be conserved at the vertex, 
hence a single summation. As before, each time such a KK threshold is 
crossed, the diagram contributes an amount identical to the SM. Then 
after summing over an infinite tower of KK modes, as shown in 
\cite{ddg}, one obtains the following simple working rule: identify the 
diagram which contains internal KK modes and multiply its SM 
contribution by $t_\delta$. In fact, $t_\delta$ represents the volume of 
a $\delta$-dimensional sphere of radius $\mu$ where the unit of volume 
is $\mu_0^\delta$ -- it counts the number of KK modes excited upto an 
energy scale $\mu$. So, in a sense, $t_\delta$ is a measure of the 
density of KK modes which accelerates the running by inducing an 
explicit $\mu^\delta$ dependence on the right hand side of 
Eqs.~(\ref{kappa}), (\ref{yukawa}) and (\ref{gauge}).  Clearly, in the 
limit $\delta = 0$, one recovers the usual logarithmic 
running. Intuitively, the power law behaviour stems from the fact that 
couplings which are dimensionless in 4 dimensions become dimensionful in 
higher dimensions. 
 
Before embarking on the main theme of the running of the neutrino 
mixing angle, we touch upon a related issue which concerns the allowed 
range of the Higgs mass. In the SM, where Higgs constitutes the only 
scalar, the requirement that the scalar potential remains bounded from 
below (i.e., $\lambda > 0$) throughout the energy thoroughfare $M_Z < 
\mu < M_{\rm GUT}$ restricts the Higgs mass to lie above $\sim 145$ 
GeV. The crucial controlling factor is, in fact, the splitting between 
the top and the Higgs masses. Supposing the Higgs to weigh $\sim 115$ 
GeV, where a preliminary hint was claimed by the LEP Collaborations, 
the one-loop RG running in the SM drives the $\lambda$ parameter 
towards negative values near a scale as close as $\sim 10^{4}-10^{5}$ 
GeV, which prompted the authors of Ref.~\cite{ellisross} to invoke the 
case for supersymmetry which prevents the occurence of a negative 
$\lambda$. In our case, which deals with only SM and its bosonic KK 
excitations, the energy interval, as we discussed before, is $M_Z < 
\mu <M$, where $M \sim 30$ TeV with $R^{-1} = 1$ TeV. We have found, 
with the RG running given by Eq.~(\ref{yukawa}), that (i) the 
stability of the potential ($\lambda > 0$) requires a lower limit $M_H 
> 98$ GeV, and (ii) the requirement of perturbativity demands an upper 
limit $M_H < 153$ GeV. Admittedly, these limits are merely indicative 
as they are based on only one loop RG evolution. 
 
Now let us take a stock of the parameters which control the running of 
$\kappa_{ij}$ and the neutrino mixing angle ($\theta$). The values of 
all the gauge and the relevant Yukawa couplings at the weak scale are 
input parameters. Similarly, a choice of the Higgs mass is necessary 
to fix the quartic coupling, $\lambda$, at the weak scale. We have 
checked that the mixing angle evolution is insensitive to the choice 
of the Higgs mass as long as the latter respects the stability and the 
perturbativity conditions of the potential. As a reference point, we 
have chosen $M_H = 115$ GeV. Then a two-step running 
determines the values of all these couplings at the scale $M$. In the 
interval $M_Z < \mu < \mu_0$, the running is logarithmic, controlled 
by the SM beta functions (putting $\delta = 0$), while in the range 
$\mu_0 < \mu < M$, power law running takes over with $\delta = 1$. We 
choose $\mu_0 = 10^3$ GeV and $M = 10^{4.5}$ GeV = 30 TeV to be our 
reference scales.  Variations of $\mu$ and $M$ around these reference 
values do not throw much insight into our agenda, and hence, for the 
sake of brevity and concise illustration, we stick to these values 
throughout this paper. The $\kappa$ matrix is defined and parametrized 
at the scale $M$ for the two-flavour case as $d\kappa \equiv 
(\kappa_{11} - \kappa_{22})/\kappa_{22}$. The other parameter to be 
fixed at $M$ is the neutrino mixing angle given by $\tan 2\theta = 
2\kappa_{12}/(\kappa_{22}-\kappa_{11})$, in a basis in which the 
charged lepton mass matrix is diagonal.  The mixing angle runs 
according to 
\begin{eqnarray} 
16\pi^2 \frac{d\sin^2 2\theta}{d\ln\mu} & = & \sin^2 2\theta 
(1 - \sin^2 2\theta) (y_2^2 - y_1^2) \frac{\kappa_{22}+\kappa_{11}} 
{\kappa_{22}-\kappa_{11}}, 
\label{angrun} 
\end{eqnarray} 
where $y_2$ and $y_1$ are the charged lepton Yukawa couplings. In our 
case, $y_2$ is $Y_\tau$ and $y_1$ is either $Y_e$ or $Y_\mu$.  
It is important to note, as emphasized by Chankowski et al. in 
\cite{rest}, that although $\theta = 0$ is a fixed point, $\theta = 
\pi/4$ is not. In fact, the evolution of $d\kappa$ 
does not have a fixed point at $d\kappa = 0$.  
 
Our goal is to choose small but non-zero values of $\sin^2 2\theta|_M$ 
and then investigate whether appropriate values of $d\kappa|_M$ exist 
which would magnify $\sin^2 2\theta|_{M_Z}$ following a two-step 
running.  An inspection of Eq.~(\ref{angrun}) reveals that this 
running would be significant only when $d\kappa$ is less than or close 
to $Y_\tau^2$. This requires $\kappa_{22} < \kappa_{11}$ at $M$. In 
fact, during the process of running, $\kappa_{11}$ and $\kappa_{22}$ 
cross each other at some energy scale leading to a resonance in the 
mixing angle at that scale. This happens due to the appearance of 
$d\kappa$ in the denominator of the right hand side of 
Eq.~(\ref{angrun}). Indeed, the scale at which this resonance occurs 
depends crucially on the interplay between $d\kappa|_M$ and the 
distinct lengths of the logarithmic and power law running determined 
by the choices of $\mu_0$ and $M$.  Our purpose is to attribute a very 
small mixing angle at $M$ and probe the appropriate parameter range 
that generates a large mixing angle near $M_Z$. 
 
\begin{floatingfigure}[r]{8cm} 
\centerline{\input{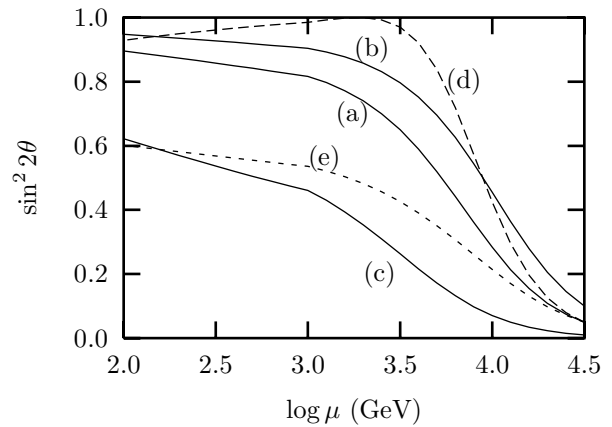}} 
  \caption[]{\small $\sin^2 2\theta$ has been plotted against the 
  renormalization scale. The values of $M_H$, $\mu_0$ and $M$ are 115 
  GeV, 1 TeV and 30 TeV, respectively. The different plots correspond 
  to different combinations of ($d\kappa|_M$, $\sin^2 2\theta|_M$) 
  given by: (a) $(1.5\times 10^{-4}, 0.05)$, (b) $(1.5\times 10^{-4}, 
  0.1)$, (c) $(1.5\times 10^{-4}, 0.01)$, (d) $(1.3\times 10^{-4}, 
  0.05)$, and (e) $(1.7\times 10^{-4}, 0.05)$. 
\label{fig1}} 
\end{floatingfigure} 
 
In Fig.~1 we have plotted $\sin^2 2\theta$ as a function of the 
renormalization scale for different values of $d\kappa|_M$ and $\sin^2 
2\theta|_M$. The graphs labelled by (a), (b) and (c) correspond to the 
choices of the initial mixing angle $\sin^2 2\theta|_M = 0.05, 0.1$ 
and $0.01$, respectively, for a fixed $d\kappa|_M = 1.5 \times 
10^{-4}$. We observe that for the plots (a) and (b) $\sin^2 
2\theta|_{M_Z}$ reaches near maximal values, while for (c) it is still 
quite large. For the other two cases (d) and (e), $\sin^2 2\theta|_M$ 
has been fixed to $0.05$, only that for (d) $d\kappa|_M = 1.3 \times 
10^{-4}$ while for (e) $d\kappa|_M = 1.7 \times 10^{-4}$.  We make 
two observations: (i) for smaller values of $d\kappa|_M$, the 
mixing angle resonance occurs at a higher scale as a result of 
$\kappa_{22} - \kappa_{11}$ approaching zero with less running 
from above, and (ii) with smaller values of $\sin^2 2\theta|_M$, 
the values of $\sin^2 2\theta|_{M_Z}$ are smaller, as expected. 
Thus, with the ultimate goal of generating a large mixing angle 
at $M_Z$ in mind, a significantly large fine-tuning is admittedly 
involved in the selection of $d\kappa|_M$, but the situation is 
not at all fine-tuned when it comes to the choice of the initial 
mixing angle.

Since $d\kappa \sim 0.5 \Delta m^2/m^2$, where $m = (m_{11} + m_{22})/2$, 
the requirement of the mixing angle resonance near $M_Z$ almost pins 
down the associated mass splitting. For the reference case $d\kappa|_M 
= 1.5 \times 10^{-4}$, we obtain $(\Delta m^2/m^2)_M = 3 \times 
10^{-4}$.  Now, we have observed that $d\kappa$ decreases by one order 
of magnitude during the RG evolution from $M$ to $M_Z$, the bulk of 
the effect coming from the power law region $M> \mu > \mu_0$. This 
means $(\Delta m^2/m^2)_{M_Z} \simeq 3 \times 10^{-5}$. According to 
the recently claimed evidence of neutrinoless double beta decay 
\cite{klap}, $m$ is expected to lie in the range 0.05 to 0.84 eV at 
95\% C.L. Now with $\nu_\tau$-$\nu_e$ oscillation in mind, with $m$ 
towards the higher end of the above range, corresponding to $\kappa 
\sim 10^{-8}$, we find a mass splitting appropriate for a MSW solar 
neutrino oscillation in the LMA region, while with $m$ sitting in the 
lower end of that range, which arises when $\kappa \sim 10^{-9}$, we 
may expect a MSW solar neutrino oscillation in the LOW region 
\cite{kamal}.  We make two observations at this point. First, it is 
not possible to produce a $\Delta m^2$ large enough to explain the 
atmospheric neutrino data. Second, if we consider $\nu_\mu$-$\nu_e$ 
oscillation, i.e., leave out $\nu_\tau$ from consideration, then 
$d\kappa$ would have to be $\sim Y_\mu^2$ to ensure mixing angle 
resonance, but the corresponding $\Delta m^2$ would be too small to 
fit any experimental data.  
 
If we take the neutrinoless double beta decay upper and lower limits
on the absolute Majorana mass seriously, then from one stand-point our
prediction can be contrasted with that of the usual 4-dimensional
model. While in our extra-dimensional case, as we pointed out, both
LMA and LOW solutions can be obtained, in the 4-dimensional scenario
only the LOW solution can be easily achieved. Interestingly, the LOW
solution is only marginally allowed after the incorporation of the SNO
neutral current data \cite{snonc}.
 
Evidently, the large mixing angle which is being sought will be in the 
so-called `dark side' if $\kappa_{22}-\kappa_{11}$ is negative and 
$\kappa_{12}$ positive or {\em vice versa}.  Only the magnitude of 
$\kappa_{12}$ is fixed by $\sin^22\theta$, while its sign is 
arbitrary. If we take $\kappa_{12}$ to be negative (positive) then the 
reference boundary value chosen, namely, $d\kappa|_M = 1.5 \times 
10^{-4}$, puts this solution in the bright (dark) side at both $M$ and 
$M_Z$ ($\sin^22\theta$ has not crossed unity in curve (a) of Fig.~1). It 
is also possible to have small mixing in the dark (bright) side at $M$ 
become large mixing in the bright (dark) side at $M_Z$ by choosing, for 
example, $d\kappa|_M = 1.3 \times 10^{-4}$ (see curve (d) in Fig.~1) and 
$\kappa_{12}$ positive (negative). 
 
The main thrust of the paper has been on the magnification of a small 
mixing angle at $M$ to a large one at the scale $M_Z$. For the examples 
that have been presented, we have verified that for the chosen 
parameters essentially {\em any} initial mixing results in a large 
mixing at the low scale.

Our main focus of attention has been the RG running of the neutrino 
mixing angles in the extra-dimensional scenario. In the process, we have 
also examined the evolution of the other SM parameters and we 
summarize the essential features now. Till the scale $\mu_0$ no KK 
modes are excited and all couplings evolve as in the SM. The running 
is different only in the $\mu_0 < \mu < M$ range. The gauge couplings, 
$g_i~(i=1,2,3)$ achieve a near equality at a scale of $1.4 \times 
10^4$ GeV, as noted already in \cite{ddg}. The quark Yukawa couplings 
run much faster than in the SM and $m_b = m_\tau$ is achieved at around 
$1.6 \times 10^4$ GeV. The evolution of the quartic scalar coupling 
$\lambda$ is critical for limiting the range of the allowed Higgs 
masses and has already been discussed earlier. Beyond $\mu_0$ it 
initially falls faster but then there is a slowing down and eventually 
even a slight increase. This is a major departure from the SM.

In summary, we have considered the effect on the RG evolution of the 
Majorana neutrino mass operator and the different SM (gauge, Yukawa, 
and quartic scalar) couplings due to the KK excitations arising from 
the compactification of one extra dimension. In the scenario under 
consideration, the fermions are restricted to a fixed brane and have 
no KK excitations, while the bosons can travel in the bulk and have 
higher KK modes. Our main conclusion is that in a two flavour picture, 
due to power law acceleration, the mixing between the $\nu_\tau$ and 
another active neutrino can achieve near maximal values at $M_Z$ even 
if it is only a few per cent at the ${\cal{O}}$ (10 TeV) scale.  It is 
worth extending our analysis to the cases which concern fermionic KK 
excitations and promoting the analysis to the study of three flavour 
oscillation. Furthermore, all these questions can be addressed in the 
context of supersymmetry.

{\bf Acknowledgements:} GB acknowledges hospitality of the CERN Theory
Division and LPT, Orsay, where part of the work was done. The research
of AR has been supported by the Council of Scientific and Industrial
Research, India.


\begin{thebibliography}{99} 
 
\bibitem{babu1} K.S. Babu, C.N. Leung, J. Pantaleone, Phys. Lett. B 
319 (1993) 191. 
 
\bibitem{chan} P.H. Chankowski, Z. Pluciennik, Phys. Lett. B 316 
(1993) 312. 
 
\bibitem{rest} P.H. Chankowski, S. Pokorski, hep-ph/0110249; 
P.H. Chankowski, W. Krolikowski, S. Pokorski, Phys. Lett. B 473 (2000) 
109; K.S. Babu, C.N. Leung, Nucl. Phys. B 619 (2001) 667; 
K.R.S. Balaji, R.N. Mohapatra, M.K. Parida, E.A. Paschos, Phys. Rev. D 
63 (2001) 113002; J. Ellis, S. Lola, Phys. Lett. B 458 (1999) 310;  
J.A. Casas, J.R. Espinosa, A. Ibarra and I. Navarro, 
Nucl. Phys. B 573 (2000) 652.    
 
\bibitem{anto} I. Antoniadis, Phys. Lett. B 246 (1990) 377; 
I. Antoniadis, K. Benakli, Phys. Lett. B 326 (1994) 69. 
 
\bibitem{ddg} K.R. Dienes, E. Dudas, T. Gherghetta, Nucl. Phys. B 537 
(1999) 47. 
 
\bibitem{addj} N. Arkani-Hamed, S. Dimopoulos, G. Dvali, 
J. March-Russell, Phys. Rev. D 65 (2002) 024032.   
 
\bibitem{bando} M. Bando, T. Kobayashi, T. Noguchi, K. Yoshioka, 
Phys. Lett. B480 (2000) 187. 
 
\bibitem{drees} S. Antusch, M. Drees, J. Kersten, M. Lindner, M. Ratz, 
Phys. Lett. B 525 (2002) 130. 
 
\bibitem{mv} M. Machacek, M.T. Vaughn, Phys. Lett. B 103 (1981) 427. 
 
\bibitem{abelking} S.A. Abel, S.F. King, Phys. Rev. D 59 (1999) 
095010. 
 
\bibitem{ellisross} J. Ellis, D.A. Ross, Phys. Lett. B 506 (2001) 
331. 
 
\bibitem{klap} H.V. Klapdor-Kleingrothaus, A. Dietz, H.L. Harney,  
I.V. Krivosheina, Mod. Phys. Lett. A 16 (2001) 2409. 
 
\bibitem{kamal} G.L. Fogli, E. Lisi, D. Montanino, A. Palazzo, 
Phys. Rev. D 64 (2001) 093007; J.N. Bahcall, M.C. Gonzalez-Garcia, 
C. Pe\~{n}a-Garay, JHEP 0108 (2001) 014; A. Bandopadhyay, S. Choubey, 
S. Goswami, K. Kar, Phys. Lett. B 519 (2001) 83; P.I. Krastev, 
A. Yu. Smirnov, hep-ph/0108177. 
 
\bibitem{snonc} Q.R. Ahmad {\em et al.}, (SNO Collaboration), 
nucl-ex/0204008. 
 
\end{thebibliography}
\end{document}